



\font\twelverm=cmr10 scaled 1200    \font\twelvei=cmmi10 scaled 1200
\font\twelvesy=cmsy10 scaled 1200   \font\twelveex=cmex10 scaled 1200
\font\twelvebf=cmbx10 scaled 1200   \font\twelvesl=cmsl10 scaled 1200
\font\twelvett=cmtt10 scaled 1200   \font\twelveit=cmti10 scaled 1200

\skewchar\twelvei='177   \skewchar\twelvesy='60


\def\twelvepoint{\normalbaselineskip=12.4pt
  \abovedisplayskip 12.4pt plus 3pt minus 9pt
  \belowdisplayskip 12.4pt plus 3pt minus 9pt
  \abovedisplayshortskip 0pt plus 3pt
  \belowdisplayshortskip 7.2pt plus 3pt minus 4pt
  \smallskipamount=3.6pt plus1.2pt minus1.2pt
  \medskipamount=7.2pt plus2.4pt minus2.4pt
  \bigskipamount=14.4pt plus4.8pt minus4.8pt
  \def\rm{\fam0\twelverm}          \def\it{\fam\itfam\twelveit}%
  \def\sl{\fam\slfam\twelvesl}     \def\bf{\fam\bffam\twelvebf}%
  \def\mit{\fam 1}                 \def\cal{\fam 2}%
  \def\tt{\twelvett}
  \textfont0=\twelverm   \scriptfont0=\tenrm   \scriptscriptfont0=\sevenrm
  \textfont1=\twelvei    \scriptfont1=\teni    \scriptscriptfont1=\seveni
  \textfont2=\twelvesy   \scriptfont2=\tensy   \scriptscriptfont2=\sevensy
  \textfont3=\twelveex   \scriptfont3=\twelveex  \scriptscriptfont3=\twelveex
  \textfont\itfam=\twelveit
  \textfont\slfam=\twelvesl
  \textfont\bffam=\twelvebf \scriptfont\bffam=\tenbf
  \scriptscriptfont\bffam=\sevenbf
  \normalbaselines\rm}



\def\beginlinemode{\endmode
  \begingroup\parskip=0pt \obeylines\def\\{\par}\def\endmode{\par\endgroup}}
\def\beginparmode{\endmode
  \begingroup \def\endmode{\par\endgroup}}
\let\endmode=\par
{\obeylines\gdef\
{}}
\def\singlespace{\baselineskip=\normalbaselineskip}

\def\oneandahalfspace{\baselineskip=\normalbaselineskip
  \multiply\baselineskip by 3 \divide\baselineskip by 2}
\def\doublespace{\baselineskip=\normalbaselineskip \multiply\baselineskip by 2}

\newcount\firstpageno
\firstpageno=-10
\footline={\ifnum\pageno<\firstpageno{\hfil}\else{\hfil\twelverm\folio\hfil}\fi}
\let\rawfootnote=\footnote		
\def\footnote#1#2{{\rm\singlespace\parindent=0pt\rawfootnote{#1}{#2}}}
\def\raggedcenter{\leftskip=4em plus 12em \rightskip=\leftskip
  \parindent=0pt \parfillskip=0pt \spaceskip=.3333em \xspaceskip=.5em
  \pretolerance=9999 \tolerance=9999
  \hyphenpenalty=9999 \exhyphenpenalty=9999 }
\def\dateline{\rightline{\ifcase\month\or
  January\or February\or March\or April\or May\or June\or
  July\or August\or September\or October\or November\or December\fi
  \space\number\year}}
\def\received{\vskip 3pt plus 0.2fill
 \centerline{\sl (Received\space\ifcase\month\or
  January\or February\or March\or April\or May\or June\or
  July\or August\or September\or October\or November\or December\fi
  \qquad, \number\year)}}


\hsize=6.5truein
\hoffset=0truein
\vsize=8.9truein
\voffset=0.0truein
\parskip=\medskipamount
\twelvepoint		
\doublespace		
\overfullrule=0pt	



\def\title			
  {\null\vskip 3pt plus 0.2fill
   \beginlinemode \doublespace \raggedcenter \bf}

\def\author			
  {\vskip 3pt plus 0.2fill \beginlinemode
   \singlespace \raggedcenter}

\def\affil			
  {\vskip 3pt plus 0.1fill \beginlinemode
   \oneandahalfspace \raggedcenter \sl}

\def\abstract			
  {\vskip 3pt plus 0.3fill \beginparmode
   \singlespace \narrower ABSTRACT: }

\def\endtitlepage		
  {\endpage			
   \body}

\def\body			
  {\beginparmode}		

\def\head#1{			
  \filbreak\vskip 0.5truein	
  {\immediate\write16{#1}
   \raggedcenter \uppercase{#1}\par}
   \nobreak\vskip 0.25truein\nobreak}

\def\refto#1{$^{#1}$}		

\def\references			
  {\head{References}		
   \beginparmode
   \frenchspacing \parindent=0pt \leftskip=1truecm
   \parskip=8pt plus 3pt \everypar{\hangindent=\parindent}}

\def\figcaps			
  {\head{Figure Captions}	
   \beginparmode
   \frenchspacing \parindent=0pt \leftskip=1truecm
   \parskip=8pt plus 3pt \everypar{\hangindent=\parindent}}

\def\publications		
  {\head{Publications}		
   \beginparmode
   \frenchspacing \parindent=0pt \leftskip=1truecm
   \parskip=8pt plus 3pt \everypar{\hangindent=\parindent}}

\gdef\refis#1{\indent\hbox to 0pt{\hss#1.~}}	

\gdef\journal#1, #2, #3, 1#4#5#6{		
    {\sl #1~}{\bf #2}, #3, (1#4#5#6)}		

\def\refstylenp{		
  \gdef\refto##1{ [##1]}				
  \gdef\refis##1{\indent\hbox to 0pt{\hss##1)~}}	
  \gdef\journal##1, ##2, ##3, ##4 {			
     {\sl ##1~}{\bf ##2~}(##3) ##4 }}

\def\refstyleprnp{		
  \gdef\refto##1{ [##1]}				
  \gdef\refis##1{\indent\hbox to 0pt{\hss##1)~}}	
  \gdef\journal##1, ##2, ##3, 1##4##5##6{		
    {\sl ##1~}{\bf ##2~}(1##4##5##6) ##3}}

\def\endreferences{\body}

\def\figurecaptions		
  {\endpage
   \beginparmode
   \head{Figure Captions}
}

\def\endpage			
  {\vfill\eject}

\def\endpaper			
  {\endmode\vfill\supereject}


\def\ref#1{Ref. #1}			
\def\Ref#1{Ref. #1}			

\def\frac#1#2{{\textstyle{#1 \over #2}}}

\def\sla{\raise.15ex\hbox{$/$}\kern-.57em}
\def\leaderfill{\leaders\hbox to 1em{\hss.\hss}\hfill}
\def\twiddle{\lower.9ex\rlap{$\kern-.1em\scriptstyle\sim$}}
\def\bigtwiddle{\lower1.ex\rlap{$\sim$}}
\def\gtwid{\mathrel{\raise.3ex\hbox{$>$\kern-.75em\lower1ex\hbox{$\sim$}}}}
\def\ltwid{\mathrel{\raise.3ex\hbox{$<$\kern-.75em\lower1ex\hbox{$\sim$}}}}
\def\square{\kern1pt\vbox{\hrule height 1.2pt\hbox{\vrule width 1.2pt\hskip 3pt
   \vbox{\vskip 6pt}\hskip 3pt\vrule width 0.6pt}\hrule height 0.6pt}\kern1pt}

\catcode`@=11
\newcount\r@fcount \r@fcount=0
\newcount\r@fcurr
\immediate\newwrite\reffile
\newif\ifr@ffile\r@ffilefalse
\def\w@rnwrite#1{\ifr@ffile\immediate\write\reffile{#1}\fi\message{#1}}

\def\writer@f#1>>{}
\def\referencefile{
  \r@ffiletrue\immediate\openout\reffile=\jobname.ref%
  \def\writer@f##1>>{\ifr@ffile\immediate\write\reffile%
    {\noexpand\refis{##1} = \csname r@fnum##1\endcsname = %
     \expandafter\expandafter\expandafter\strip@t\expandafter%
     \meaning\csname r@ftext\csname r@fnum##1\endcsname\endcsname}\fi}%
  \def\strip@t##1>>{}}

\def\citeall#1{\xdef#1##1{#1{\noexpand\cite{##1}}}}
\def\cite#1{\each@rg\citer@nge{#1}}	

\def\each@rg#1#2{{\let\thecsname=#1\expandafter\first@rg#2,\end,}}
\def\first@rg#1,{\thecsname{#1}\apply@rg}	
\def\apply@rg#1,{\ifx\end#1\let\next=\relax
\else,\thecsname{#1}\let\next=\apply@rg\fi\next}

\def\citer@nge#1{\citedor@nge#1-\end-}	
\def\citer@ngeat#1\end-{#1}
\def\citedor@nge#1-#2-{\ifx\end#2\r@featspace#1 
  \else\citel@@p{#1}{#2}\citer@ngeat\fi}	
\def\citel@@p#1#2{\ifnum#1>#2{\errmessage{Reference range #1-#2\space is bad.}
    \errhelp{If you cite a series of references by the notation M-N, then M and
    N must be integers, and N must be greater than or equal to M.}}\else%
 {\count0=#1\count1=#2\advance\count1
by1\relax\expandafter\r@fcite\the\count0,%
  \loop\advance\count0 by1\relax
    \ifnum\count0<\count1,\expandafter\r@fcite\the\count0,%
  \repeat}\fi}

\def\r@featspace#1#2 {\r@fcite#1#2,}	
\def\r@fcite#1,{\ifuncit@d{#1}		
    \expandafter\gdef\csname r@ftext\number\r@fcount\endcsname%
    {\message{Reference #1 to be supplied.}\writer@f#1>>#1 to be supplied.\par
     }\fi%
  \csname r@fnum#1\endcsname}

\def\ifuncit@d#1{\expandafter\ifx\csname r@fnum#1\endcsname\relax%
\global\advance\r@fcount by1%
\expandafter\xdef\csname r@fnum#1\endcsname{\number\r@fcount}}

\let\r@fis=\refis			
\def\refis#1#2#3\par{\ifuncit@d{#1}
    \w@rnwrite{Reference #1=\number\r@fcount\space is not cited up to now.}\fi%
  \expandafter\gdef\csname r@ftext\csname r@fnum#1\endcsname\endcsname%
  {\writer@f#1>>#2#3\par}}

\def\r@ferr{\endreferences\errmessage{I was expecting to see
\noexpand\endreferences before now;  I have inserted it here.}}
\let\r@ferences=\references
\def\references{\r@ferences\def\endmode{\r@ferr\par\endgroup}}

\let\endr@ferences=\endreferences
\def\endreferences{\r@fcurr=0
  {\loop\ifnum\r@fcurr<\r@fcount
    \advance\r@fcurr by 1\relax\expandafter\r@fis\expandafter{\number\r@fcurr}%
    \csname r@ftext\number\r@fcurr\endcsname%
  \repeat}\gdef\r@ferr{}\endr@ferences}


\let\r@fend=\endpaper\gdef\endpaper{\ifr@ffile
\immediate\write16{Cross References written on []\jobname.REF.}\fi\r@fend}

\catcode`@=12

\citeall\refto		
\citeall\ref		%
\citeall\Ref		%

\def\invisiblerefto#1{\setbox111=\hbox{#1}}
\citeall\invisiblerefto


\title{\bf The fermion sign problem: \\
A new decoupling transformation, and a new simulation algorithm}

\vskip1.0cm

\author Ghassan George Batrouni$^a$ and Philippe de Forcrand$^b$$^\dagger$

\affil
(a) Thinking Machines Corporation
245 First Street
Cambridge, MA 02142
ggb@think.com

\vskip0.3cm
\affil
(b) Theoretical Physics Institute
University of Minnesota
Minneapolis, MN 55455
forcrand@ips.id.ethz.ch

\abstract
\singlespace
We discuss the Fermion sign problem and, by examining a very general
Hubbard-Stratonovich (HS) transformation, argue that the sign problem
cannot be solved with such methods. We propose a different kind of
transformation which, while not solving the sign problem,
shows more detailed information about the system.
With our transformation it is {\it trivial} to tell which
auxiliary field configurations give a positive sign and which give a negative
sign. We then discuss briefly various properties of this transformation and
construct a new algorithm which with one simulation gives results for a
whole range of particle densities and Hubbard $U$ values, positive and
negative. Our approach is in excellent agreement with exact calculations.

\vskip 1in
\vskip 1.2in
$\dagger$  Permanent address: IPS, ETHZ, CH-8092 Zurich Switzerland.
\vfill
\endtitlepage

\doublespace
The major obstacle facing numerical simulation of a large number of
strongly interacting electrons is the so-called ``fermion sign
problem''.\refto{loh,sorella,white,hirsch,batrouni,tremblay,assaad}
This problem appears in many different guises and here we will
focus on its form in the determinant algorithm for quantum Monte Carlo. Our
method and conclusions are general but for clarity we will
concentrate on the Hubbard model, which is under active study as a
model for metal-insulator transitions and high temperature superconductivity.
We begin by reviewing the Hubbard-Stratonovich transformation and the
resulting sign problem.
The partition function of the Hubbard model in the grand canonical ensemble
is given by
$$Z = tr\bigl (e^{-\beta H} \bigr ),\eqno(1)$$
$$\eqalignno{H
&= -t\sum_{<ij>,\sigma} \bigl( c^{\dagger}_{\sigma}(i) c_{\sigma}^{}(j) +
c^{\dagger}_{\sigma}(j)c_{\sigma}^{}(i)\bigr) + U\sum_{i}\bigl(
n_{+}^{}(i)-{1\over2} \bigr)\bigl( n_{-}^{}(i)-{1\over2} \bigr)\cr
&\qquad-\mu \sum_{i} \bigl( n_{+}^{}(i)+n_{-}^{}(i) \bigr),&(2a)\cr
&= \sum_{ij,\sigma} c^{\dagger}_{\sigma}(i)k_{ij}c_{\sigma}(j)
+U\sum_{i}\bigl(
n_{+}^{}(i)-{1\over2} \bigr)\bigl( n_{-}^{}(i)-{1\over2} \bigr).&(2b)\cr}$$
The
sum $<ij>$ is over all pairs of nearest neighbor lattice sites, $t$ is the
hopping parameter, $c^{\dagger}_{\sigma}(i)$ and $c_{\sigma}(i)$ are
creation and annihilation operators of electrons with spin $\sigma$ along
the $z$ axis at site $i$. $n_{+}$ ($n_{-}$) is the number
operator for electrons with up (down) spins, $\beta$ is the inverse
temperature, and $U$ is the coupling constant, which can be
positive or negative. The matrix $k$ in eq.(2b) contains the hopping term
and the chemical potential. Through standard techniques one can transform the
trace into a path integral (or sum) over the configurations of a c-number
auxiliary field.\refto{blanken} First we use the Trotter-Suzuki
approximation\refto{trotter} to express $Z$
as: $Z=tr(e^{-\tau H})^{L} \approx tr(e^{-\tau k}e^{-\tau
V})^{L}$, where $\tau = {\beta \over L}\ll 1$, is the imaginary time step,
and $L$ is the number of such time steps. Now we use the HS transformation
to decouple the quartic potential term, $e^{-\tau V}$, into quadratics in
the creation and annihilation operators:
$$e^{-\tau U\bigl (n_{+}^{}(i)-{1\over2}\bigr
)\bigl (n_{-}^{}(i)-{1\over2}
\bigr)} = {{e^{-\tau U/4}}\over2} \sum_{s(i,l)=\pm 1}
e^{-\lambda s(i,l) \bigl (n_{+}^{}(i) - n_{-}^{}(i)\bigr )}\eqno(3)$$
at each site $i$ and time slice $l$. $\lambda$ is related to $U$ via
$cosh(\lambda)=e^{{\tau U}\over 2}$.
Here we wrote the discrete HS
transformation,\refto{hirsch} but one can also write a continuous one.
The trace over the fermion operators can now be taken since they only appear
quadratically. This gives:\refto{blanken}
$$Z=\sum_{s(i,l)=\pm 1} detM^{+} detM^{-},\eqno(4)$$
where
$$M^{\sigma}=I+ B^{\sigma}_{L}B^{\sigma}_{L-1} ...B^{\sigma}_{1},\eqno(5)$$
and
$$B^{\pm}_{l}=e^{\mp\lambda v(l)} e^{-\tau k}.\eqno(6)$$
$I$ is a $V\times V$ unit matrix, $V$ is the spatial volume,
$v(l)_{ij}=\delta_{ij}s(i,l)$, where i runs from $1$ to $V$ and $l$ from
$1$ to $L$. The partition function is
now written as a sum over c-numbers and can therefore be simulated on a
computer. The sign problem arises because the determinants and their
product, in eq(4), can be negative and thus cannot be used
as the probability density in a Monte Carlo simulation.
Instead, their absolute value is used, and the average
of an observable, $A$, is
then given by $<A>=<Asgn>^{\prime}/<sgn>^{\prime}$, where $<>^{\prime}$ denotes
averages with respect to the absolute value of the determinants. The
average sign is thus defined as $<sgn>^{\prime}=Z/Z^{\prime}$, where
$Z^{\prime}$ is the
partition function resulting from using the absolute value of the
determinants. This approach works well for small $U$ and relatively high
temperatures. As the temperature decreases ($\beta$ increases), the
average sign scales like\refto{white,batrouni} $<sgn> \sim e^{-c\beta}$,
where $c$ is a constant. This
makes low temperature simulations impractical because averages are obtained
as the ratios of two very small numbers, each with a very large variance.
When $U<0$, a similar procedure yields $M^{+} =M^{-}$. Consequently, the
product of the two determinants in eq (4) becomes a square, and therefore
positive semidefinite even though the determinant itself is still {\it
not} positive semidefinite. Thus, there is no sign problem for the
negative $U$ Hubbard model (or other attractive interactions).

The HS transformation discussed above is the most commonly used one, but it
is only one of an infinite number of possible transformations. For example,
other decoupling schemes were discussed in Refs.($5,6$)
in the hope of finding a transformation which will
solve the sign problem or at least decrease its severity. We will argue
here that there exists no HS transformation that will
eliminate the sign problem. We start by noting that the purpose of
any HS transformation is to decouple the quartic fermionic interaction into
quadratic terms which are coupled to the HS (auxiliary) field. This allows
us to perform the trace in the partition function, giving two determinants.
But,
except for minor details, all transformations examined so far have yielded
either a product of different determinants of the above form, or a single
determinant\refto{batrouni} which have always suffered severely from
the sign problem for certain values of $\mu$, $\beta$ and $U$.
This suggests that one way to solve the sign problem in this approach is
to obtain a square of a determinant. Is it, therefore, possible to
generalize the above HS
transformations such that the resulting partition function is a sum (or
integral) over a  square, and if not, why? In order to get $(detM)^2$,
the relative minus sign between $n_{-}$ and $n_{+}$ on the right hand side
of eq (3) must become a plus, thus preserving the up-down
symmetry of the Hamiltonian in the HS transformation.  Let us therefore
propose a general HS transformation
$$e^{-\tau U(n_{+}n_{-} - {1\over 2}n_{+}- {1\over 2}n_{-})}
= \int dy P(y) y^{(n_{+} + n_{-})},\eqno(7)$$
where $P(y)$ and $y$ are real\refto{transformation} and arbitrary except
for the constraints discussed below, and $U$ is positive or negative.
This transformation includes discrete transformations like eq (3), but is more
general. If
such a transformation were possible, the sign problem would be solved because
of the resulting $(detM)^{2}$. Since $n_{\pm}=0,1$, we see that the
conditions on $P(y)$ and $y$ are
$$\int dy P(y) = 1,\eqno(8)$$
$$<y>= e^{ {\tau U \over 2}},\eqno(9)$$
$$<y^{2}> = 1,\eqno(10)$$
where $<>$ means an average with respect to the weight $P(y)$. In general,
the inequality $<y>^2 \leq <y^{2}>$ must be satisfied. Combining this with
eqs (9,10) forces $U$ to be negative: In other words, the inequality
cannot be satisfied for positive values of $U$. Therefore, there is no
general HS transformation that is capable of giving the square of a
determinant. Consequently, this approach to solving the sign problem fails.

Implicit in the above argument is the positivity of $P(y)$. We will
reserve the name ``Hubbard-Stratonovich transformation'' for these cases
since all previous applications of the $HS$ transformation assume such
positivity. However, this argument is invalid if we allow
$P(y)$ to take negative values.  This of course would {\it not} solve the sign
problem, but it would give us a square of a determinant, with the result
that the sign changes now come not from the determinants, but from $P(y)$.
One advantage of this is that contrary to the HS transformation where,
in general, we do not
know which auxiliary field configurations lead to minus signs because the
structure of the determinants is too complicated, here the minus sign comes
from $P(y)$ which we know exactly. Therefore, we have complete prior
knowledge of the sign of all the configurations.

For example, a simple choice is $$P(y)= \prod_{i,l}\Bigl( a\delta
(y(i,l)-y_{1}) + b\delta(y(i,l)-y_{2})\Bigr),\eqno(11)$$
where $i$ is the space index, and $l$ is the time slice index. $y_1$ and
$y_2$ are the allowed values for the auxiliary field $y$,
$a$ and $b$ are parameters to be determined from the constraints
eqs(8,9,10). We chose two discrete values, $y_1$ and $y_2$,
for the auxiliary field, but
we could have chosen any number of discrete variables, or any
continuous distribution as long as we satisfy conditions eqs(8,9,10).
Our motivation is simplicity.
Applying the above constraints gives $a+b=1$ and
$$b= { {y_{1}^{2} -1}\over {y_{1}^{2} - y_{2}^{2}}},\eqno(12)$$
$$e^{ {\tau U}\over 2} = { {1 + y_1 y_2} \over {y_1+y_2}}.\eqno(13)$$
Having chosen the form of $P(y)$, we still have the freedom of choosing
the values of one of the two parameters $y_1$, $y_2$, the second being
determined by eq(13).

Now that we have decoupled the quartic fermion interaction into two
quadratic terms with the same sign, the trace over the fermi operators can
be performed as in the HS case giving for the partition function
$$\eqalignno{Z &= \sum_{y(i,l)=y_1,y_2} a^{n_1}b^{n_2} (detM)^2&(14)\cr
&= a^N \sum_{n_1} C^{N}_{n_2}\Bigl ({b \over a}\Bigr )^{n_2}
<(detM)^2>_{n_2}.&(15)\cr}$$
$M$ has the same form as $M^{-}$, eq(5), and
$$B_l = v(l)e^{-\tau k},\eqno(16)$$
with $v(l)_{i,j}=\delta_{ij}y(i,l)$. $n_1$ ($n_2$) is the
number of auxiliary spins with the value $y_1$ ($y_2$), and $N=n_1+n_2$ is
the total number of sites on the $(d+1)$ dimensional lattice. $<>_{n_2}$ is
an average over all
configurations which have $n_2$ spins equal to $y_2$, whose number is
the binomial coefficient $C^{N}_{n_2}$.
It is easy to show that for $U<0$, both $a$ and
$b$ are positive and therefore there is no sign problem, just like the
usual HS transformations. When $U>0$, $a$ and $b$ have {\it opposite}
signs (we take $b<0$) and thus the sign problem
reappears. However, although the value of $(detM)^2$ (always
positive) depends on {\it both} the relative number of spins with values
$y_1$ and $y_2$, and their configuration on the lattice, the prefactor
$a^{n_1}b^{n_2}$ depends {\it only} on the relative numbers. In particular,
since the source of the sign problem in our formulation is the opposite
sign of $a$ and $b$, we have complete knowledge of all the configurations
that change the sign: only configurations with an odd number of $y_2$ spins
lead to an odd exponent for $b$ and thus a negative contribution. This
complete characterization of the negative configurations is to be contrasted
with all the HS transformations previously used where one knows very little
about the kind of configurations that lead to minus signs. This vividly
demonstrates, yet again,\refto{batrouni} that the sign problem in the
determinant
algorithm is not related to configurations where the electron ``paths''
exchange. It is merely an artifact of the transformation used to decouple
the quartic terms in the Hamiltonian. Another property of our
transformation is that it preserves the rotational spin symmetry of the
original Hamiltonian.  We can therefore take measurements along any spin
direction, or along all three, thus reducing fluctuations.

Not surprisingly, the behaviour of the average sign and the summand in
eqs (14,15) depend on the values we choose for the auxiliary fields. Three
choices are of particular interest to us. The first is to choose
$y_1\approx y_2$. This choice is remarkable because the entire phase space
is explored with a variable that fluctuates very little. Consequently
the values of the determinant and other observables fluctuate little, and
change very smoothly as more spins are flipped. The disadvantage of this
choice is that $a/b \rightarrow -1$ (always keeping $a+b=1$) as $y_1
\rightarrow y_2$, which means that the contributions of the negative
configurations are of the same size as the positive ones. This makes
numerical simulations hard, but may offer
the possibility of studying the system semianalytically.

Our second choice is to take $y_2=0$. What is intriguing about this choice
is that when enough auxiliary spins have the value $y_2=0$, there will be
many realizations where the auxiliary field for at least one entire time
slice is zero, preventing the percolation of $y_1$ spins and resulting in
$M=I$ and $detM=1$. When this happens the
observables will also have a trivial value. So, when such configurations
become important, there will be many instances where the observables are
trivial and it would be interesting to study the effect such configurations
might have on phase transitions.

The third choice is to take $y=\pm1$. The limit of eqs (12,13) as
$y_1\rightarrow +1$ and $y_2\rightarrow -1$ is
$$a={1 \over 2}(1+e^{{\tau U}\over 2}),\eqno(17)$$
$$b={1 \over 2}(1-e^{{\tau U}\over 2}).\eqno(18)$$
Notice that with this choice, $M$ is {\it no longer} a function of the
coupling constant $U$, since the coupling constant can appear only in $y$,
$a$ and $b$, and here we fixed $y=\pm1$. Furthermore, recall that, in the
matrix $M$, $\mu$ appears in the form $e^{\tau\mu} I$ for each
$B^{\sigma}_{l}$ matrix, where $I$ is the $V\times V$ identity.
Consequently, the matrix $M^{\prime}=e^{-\beta \mu}(M-I)$ is a function of
temperature, but {\it not} $U$ or $\mu$. This major simplification allows
us to collect a large number of realizations of $M^{\prime}$, and then
perform data analysis for any $U$ and $\mu$. The $\mu$ dependence
is obtained by multiplying $M^{\prime}$ by $e^{\beta \mu}$. The $U$ dependence
appears through the ratio $b / a$, eqs.(15,17,18), in a way similar to
perturbation theory for $\tau$ small. We can thus obtain results for a wide
range of values of $U$ (positive and negative) and $\mu$ from only one computer
run. Currently, we generate the realizations of $M^{\prime}$ as follows: start
with all $y=+1$ and successively flip randomly selected spins to $y=-1$,
calculating $M^{\prime}$ after each spin flip. This is done up to some
maximum number of flipped spins, much smaller than the total number of sites.
The reason is that the expansion of Z in $n_2$ (eq.15), which is similar to
a perturbation expansion for $\tau$ small, converges quickly and can be
truncated. Then we start all over, and in this way we
generate an ensemble of realizations of flipped spins over which we can
average. The addition of importance sampling could greatly increase the
efficiency of the algorithm.

We tested this algorithm on a $2\times 2$ lattice and compared with exact
results. In fig.(1) we show a plot of $<n_{\uparrow}>$ versus $U$, with the
crosses showing exact results. Fig.(2) shows a similar figure for the
ferromagnetic correlation function, $S(0,0)$. We used $256$ time slices
in order to eliminate the finite time step errors for comparison with exact
diagonalization. In general, the finite $\tau$ errors are $O(\tau^2 U)$, as
in the usual determinant algorithm.
Note that as we move away
from half filling the errors increase appreciably for positive $U$ because
the sign problem is appearing in force. We also see that for
$U<0$, where there is no sign problem, errors are very small.
Recall that all of the shown curves were obtained from the data of only
one run, and that those data contain all the information needed to measure all
the equal time correlation functions for positive and negative $U$ and a
wide range of $\mu$. One way to get better results for $U>0$ away from half
filling is to increase statistics. Another more efficient way is to include
importance sampling, and to correlate the update of positive and negative
configurations since we know their signs. This would not change the average
sign but would greatly decrease its variance and that of all measurements
\refto{assaad}. This work is in progress.

\vskip0.3cm
\centerline{$\underline{\bf {\rm Acknowledgements}}$}
\vskip0.2cm
We would like to thank A. Krasnitz and R. Scalettar for very helpful
discussions and  (R.S.) for
providing exact diagonalization results against which we compared our
algorithm. GGB thanks the Theoretical Physics Institute at the University
of Minnesota for its hospitality when this work was in its early stages.

\references


\refis{loh}
E. Y. Loh Jr, J. E. Gubernatis, R. T. Scalettar, S. R. White, D. J.
Scalapino, and R. L. Sugar, Phys. Rev. {\bf B41}, 9301 (1990).

\refis{sorella}
S. Sorella, S. Baroni, R. Car, and M. Parinello, Europhysics Letters {\bf 8}
663 (1989);
S. Sorella, E. Tosatti, S. Baroni, R. Car, and M. Parinello, Int. J. Mod. Phys.
{\bf B1},
993 (1988).

\refis{white}
S. R. White, D. J. Scalapino, R. L. Sugar, E. Y. Loh, J. E. Gubernatis, and R.
T.
Scalettar, Phys. Rev. {\bf B40}, 506 (1989).

\refis{hirsch}
J. E. Hirsch, Phys. Rev. {\bf B31}, 4403 (1985).

\refis{batrouni}
G. G. Batrouni and R. T. Scalettar, {\it Physical Review} {\bf B}42, (1990)
2282.

\refis{tremblay} Liang Chen and A.-M.S. Tremblay,
Int. J. Mod. Phys. {\bf B6} (1992) 547.

\refis{blanken}
R. Blankenbecler, D. J. Scalapino, and R. L. Sugar, Phys. Rev. {\bf D24}, 2278
(1981).

\refis{trotter}
H. F. Trotter, Proc. Am. Math. Soc. {\bf 10}, 545 (1959); M. Suzuki, Comm.
Math. Phys.
{\bf 51}, 183 (1976).

\refis{transformation}
This can be easily generalized to complex variables and $det(M^{\dag} M)$,
but to keep the notation simple we use real values.

\refis{assaad}
Antithetic Langevin variables : a way to take care of the fermionic sign
problem ?, F.F. Assaad and Ph. de Forcrand, in "Quantum simulations of
condensed matter phenomena", p.1-11 (World Scientific Pub., 1990).

\endreferences

\vfill
\eject

\centerline{\bf {\rm Figure Captions}}

Figure 1. The average density, $<n_{\uparrow}>$, vs. $U$ for
$\mu=-0.1,-0.2,-0.3,-0.5$ and $-1$ as labeled in the figure. The crosses
show exact results.

Figure 2. The ferromagnetic correlation function, $S(0,0)$, vs. $U$ for the
same values of the chemical potential as in figure 1. The crosses show
exact results.

\end